\documentclass[
aps,%
12pt,%
final,%
notitlepage,%
oneside,%
onecolumn,%
nobibnotes,%
nofootinbib,%
superscriptaddress,%
noshowpacs,%
centertags]%
{revtex4}
\usepackage {amsmath}
\usepackage{graphicx,epsfig}

\begin{document}

\title{Dynamics of Vortex Pair in Radial Flow}

\author{\firstname{E.~Yu.}~\surname{Bannikova}}

\email{bannikova@astron.kharkov.ua}
\affiliation{%
Institute of Radio Astronomy, National Academy of Sciences of
Ukraine, Kharkov, 61002 Ukraine}
\affiliation{Karazin National
University, Kharkov, 61077 Ukraine}

\author{\firstname{V.~M.}~\surname{Kontorovich}}
\email{vkont@ri.kharkov.ua}
\affiliation{%
Institute of Radio Astronomy, National Academy of Sciences of
Ukraine, Kharkov, 61002 Ukraine} \affiliation{Karazin National
University, Kharkov, 61077 Ukraine}

\author{\firstname{G.~M.}~\surname{Reznik}}
\email{greznik11@yahoo.com}
\affiliation{%
Shirshov Institute of Oceanology, Russian Academy of Sciences,
Moscow, 117997 Russia}%

\begin{abstract}
\vspace {0.5cm} The problem of vortex pair motion in
two-dimensional plane radial flow is solved. Under certain
conditions for flow parameters, the vortex pair can reverse its
motion within a bounded region. The vortex-pair translational
velocity decreases or increases after passing through the
source/sink region, depending on whether the flow is diverging or
converging, respectively. The rotational motion of two corotating
vortexes in a quiescent environment transforms into motion along a
logarithmic spiral in the presence of radial flow. The problem may
have applications in astrophysics and geophysics.
\end{abstract}

\maketitle

\section{INTRODUCTION}

In the number of astrophysical and geophysical problems the
situation arises in which a system of vortices (in the simplest
case, a vortex pair) interacts with diverging or converging radial
flow. For example, in the dipolar vortex model of an active
galactic nucleus (AGN) [1], the obscuring tori\footnote{The idea
of obscuring tori underlies the unified model of AGN [3, 4] in
which the difference between types of active nuclei, such as radio
galaxies and quasars (as well as between types of Seyfert
galaxies), is attributed to the position of the obscuring tori
relative to the line of sight. For example, in the case of quasars
is possible to observe the central AGN region adjacent to the
black hole at the center of a galaxy, whereas the central region
of a radio galaxy is obscured by the surface of a torus. In 2004,
the idea was substantiated by direct observations of the obscuring
torus with an optical interferometer combining 8.2 m telescopes at
the Southern European Observatory [5].}  [2 -- 5] are toroidal
vortices in which selfgravitation is balanced by centrifugal
forces [6]. (Possible geophysical applications are discussed at
the end of this paper.)

The vortex motion inside an obscuring torus transform it from a
pure geometrical object to the dynamical one that allows to
describe dynamical processes in AGNs. The vortex motion can be
caused by interaction between the torus and the radial outflow of
wind and radiation from the AGN central region, which is
responsible also for the dipolar structure of the vortex motion in
the torus (see Fig. 1).

\begin{figure}
\includegraphics[scale=0.8]{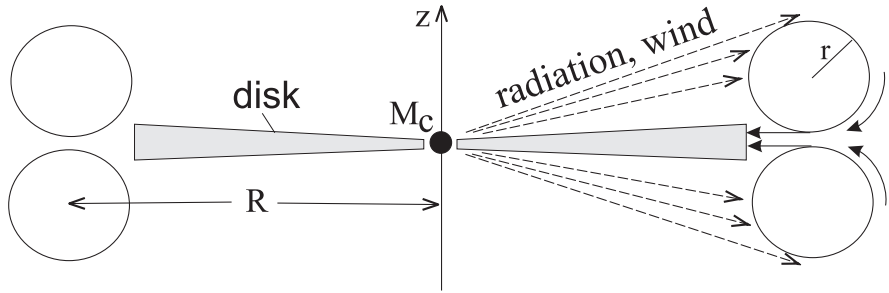}
\caption{Dipolar vortex model of AGN obscuring tori (symmetry
plane cross section) [1]: $M_c$ is the central mass and $R$ is the
major radius of a torus.}
\end{figure}

The foregoing discussion motivates a study of the dynamics of
interaction between vortices and radial flow. Problems of this
kind seem to have never been analyzed. The simplest model  is the
planar interaction between radial flow and a dipole vortex pair.
(The corresponding flow pattern can be envisaged from Fig. 1, but
the present analysis is restricted to a plane two-dimensional
problem.) In this paper, we present an exact solution for the
vortex motion in an inviscid fluid. Possible exactly solvable
generalization for rotating radial flow is considered at the end
of the paper.

\section{PLANAR VORTICES IN DIVERGING RADIAL FLOW}
The input time-dependent streamfunction equation in the plane case
has the form [7]
\begin{equation}\label{eq1}
  \frac{\partial\triangle\psi}{\partial t} + J\left( \psi , \triangle
  \psi\right) = 0,
\end{equation}
where the flow velocity {\bf v} is related to the streamfunction
$\psi$ as
\[
\text{v}_x = - \frac{\partial\psi}{\partial y} , \qquad \text{v}_y
= \frac{\partial\psi}{\partial x},  \qquad
\triangle\psi=\text{curl}_z \textbf{v}~.
\]
and $J (\alpha, \beta)$ is the Jacobian determinant. The equation
(1) is the $z$ component of the vorticity conservation equation in
two-dimensional case
\begin{equation}\label{eq2}
\frac{d}{d t}\text{curl}_z {\bf v} = 0, \qquad \frac{d}{dt} \equiv
\frac{\partial}{\partial t} + {\bf v}\cdot \nabla~.
\end{equation}

We represent the streamfunction as a decomposition into a regular
part $\psi_r$ describing the flow and a singular part $\psi_s$
whose components describe point vortices:
\begin{equation}\label{eq3}
  \psi = \psi_r + \psi_s~,
\end{equation}
where
\[
  \psi_s = \frac{1}{2\pi}\sum_m A_m \ln|{\bf r} - {\bf r}_m|~,
\]
$A_m$ is the $m$th vortex strength, and $r_m$ is the corresponding
position vector.\footnote{We use the approach developed in [8].}

The singular component satisfies the Poisson Equation [9, 10]
\begin{equation}\label{eq4}
  \triangle \psi_s = \sum_m A_m \delta(x - x_m)\delta(y - y_m).
\end{equation}
Substituting (3) into (1) we obtain the streamfunction equation
\begin{equation}\label{eq5}
  \frac{\partial \triangle\psi_r}{\partial t} +
  J\left(\psi_s + \psi_r, \triangle\psi_r\right) = 0
\end{equation}
and expressions for the velocity components of the $m$th vortex
\begin{equation}\label{eq6}
\begin{split}
 \dot{x}_m=- \left.\frac{\partial(\psi_r + \psi_s^m)}
                        {\partial y}\right|_{{\bf r}={\bf
                        r}_m}~,\\
  \dot{y}_m= \left.\frac{\partial(\psi_r + \psi_s^m)}
                        {\partial x}\right|_{{\bf r}={\bf r}_m},
\end{split}
\end{equation}
where $\psi_s^m$ is the streamfunction $\psi_s$ minus the
contribution of the $m$th vortex.

If the regular component $\psi_r$ corresponds to a diverging
radial flow with source at the origin,\footnote{Note that the
assumption of incompressible flow is inapplicable within a small
neighborhood of the radial flow source depending on sound
velocity.}
\begin{equation}\label{eq7}
\psi_r = -Q\varphi, \qquad  Q = const>0
\end{equation}
then vorticity vanishes outside the source:
\begin{equation}\label{eq8}
  \triangle \psi_r = \frac{\partial^2 \psi_r}{\partial r^2} +
                     \frac{1}{r}\frac{\partial \psi_r}{\partial r}
                     + \frac{1}{r^2}\frac{\partial^2 \psi_r}{\partial
                     \varphi^2}= 0~.
\end{equation}
Now, Eq.(5) becomes an identity, and expressions (6) combined with
(7) yield
\begin{equation}\label{eq9}
\begin{split}
 \dot{x}_m=- \left.\frac{\partial\psi_s^m}
                        {\partial y}\right|_{{\bf r}=
                        {\bf r}_m} + Q \frac{x_m}{r_m^2}~, \\
  \dot{y}_m= \left.\frac{\partial\psi_s^m}
                        {\partial x}\right|_{{\bf r}={\bf r}_m} + Q
                        \frac{y_m}{r_m^2}~.
\end{split}
\end{equation}
In the case of two vortices (Fig. 2), Eqs. (9) reduce to
\begin{equation}\label{eq10}
\begin{split}
 \dot{x}_1=- \frac{A_2}{2\pi}\frac{y_{12}}{r_{12}^2} + Q\frac{x_{1}}{r_{1}^2}~, \qquad
 \dot{y}_1= \frac{A_2}{2\pi}\frac{x_{12}}{r_{12}^2} + Q\frac{y_{1}}{r_{1}^2}~,\\
 \dot{x}_2= \frac{A_1}{2\pi}\frac{y_{12}}{r_{12}^2} + Q\frac{x_{2}}{r_{2}^2}~, \qquad
 \dot{y}_2=- \frac{A_1}{2\pi}\frac{x_{12}}{r_{12}^2} + Q\frac{y_{2}}{r_{2}^2}~.
\end{split}
\end{equation}
where $(x_{12}, y_{12} ) = (x_1 - x_2, y_1 - y_2 )$ and $r_{12}=
|{\bf r}_1 - {\bf r}_2 |$.
\begin{figure}[bth]
\centering \includegraphics[width=7cm]{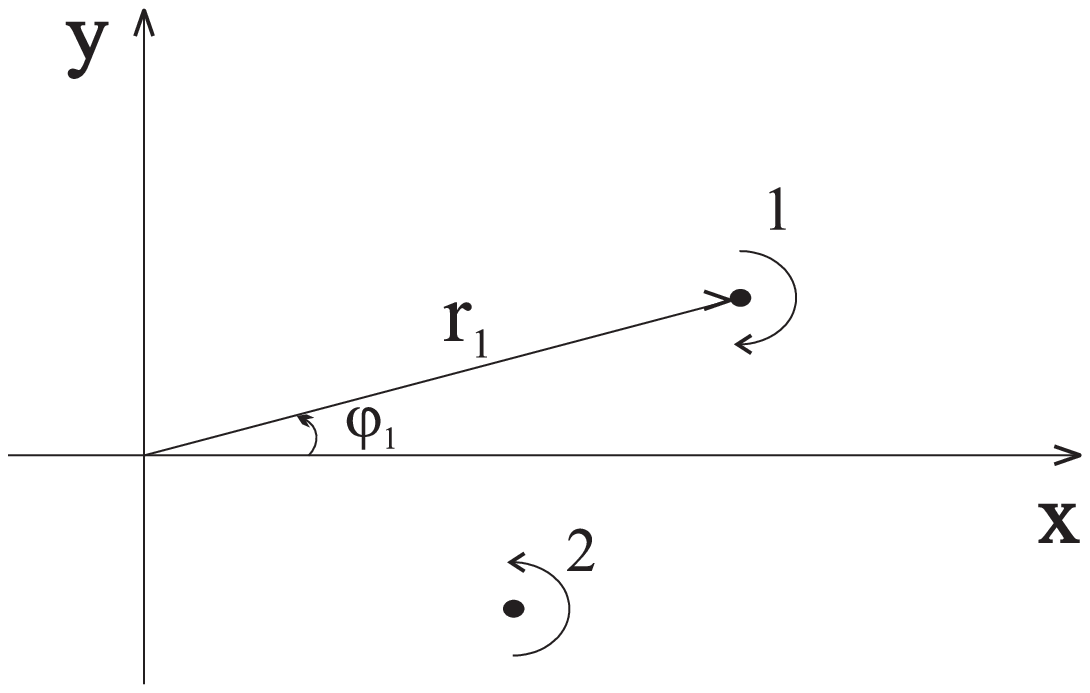} \hfill
\includegraphics
[width=7cm]{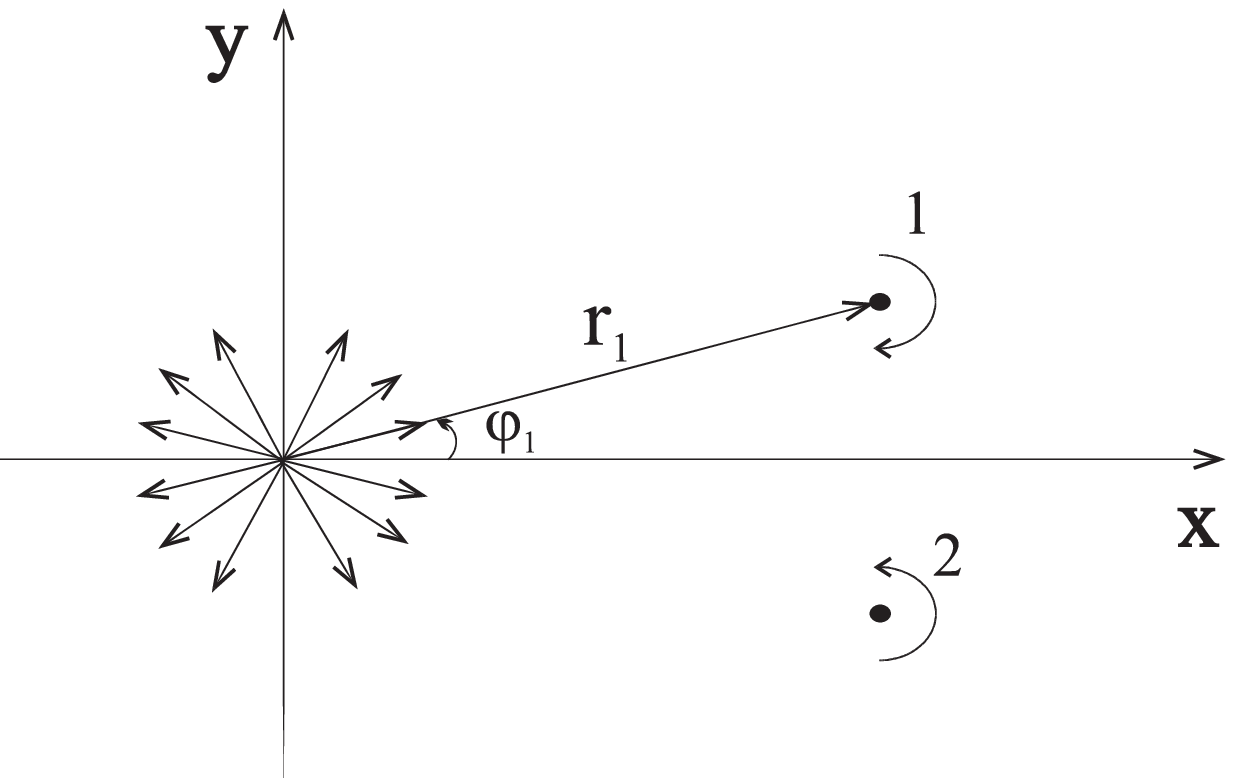}
\newline
\parbox[t]{7cm}
{\caption{Two point vortices, 1 and 2, in Cartesian and polar
coordinates. Arrows directions correspond to $A_1 < 0$ and $A_2>~
0$.}\label{fig2}} \hfill\parbox[t]{7cm}{\caption{Vortex pair in
diverging radial flow.}\label{fig3}}
\end{figure}
Consider a dipole vortex pair $(A_2 = - A_1)$ symmetric about the
$x$ axis (Fig.3) in the initial moment of time. Since $x_{12} = 0$
and $y_2 = -y_1$, it will suffice to analyze the equations of
motion of the one vortex located at $y = y_1 > 0$:
\begin{subequations}
\label{eq11}
\begin{equation}
\dot{x} =- \frac{A}{4\pi y} + Q\frac{x}{r^2}~,\label{eq11-a}
\end{equation}
\begin{equation}
 \dot{y} =Q\frac{y}{r^2}~, \label{eq11-b}
\end{equation}
\end{subequations}
where $x = x_1 = x_2$, $r = \sqrt{x^2 + y^2}$, and $A = A_2 > 0$.
In a quiescent environment, the pair moves  with velocity $A/4\pi
y$ antiparallel to the $x$ axis. It follows from Eq.(11b) that
$\dot{y} > 0$ i.e., the separation between the vortices increases
with time elapsed, while the interaction between them (the first
term in Eq.(11a)) decreases in a diverging flow.

\section{HAMILTONIAN FORMULATION AND INTEGRATION OF GOVERNING EQUATIONS}

An arbitrary system of vortices satisfying Eqs.(9) can be
described in terms of the Hamiltonian
\[
 \widehat{H} = \frac{1}{4\pi}\sum_{m \neq n} A_m A_n \ln r_{mn} -
      Q \sum_{m} A_m \text{arccot}\frac{x_m}{y_m}~.
\]
Indeed, multiplying Eqs.(9) by $A_m$ and verifying that
\[
\dot{x}_m \frac{\partial \widehat{H}}{\partial x_m} + \dot{y}_m
\frac{\partial \widehat{H}}{\partial y_m} = 0, \qquad m = 1,2,...,
\]
we can write
\[
A_m\dot{x}_m = -\frac{\partial \widehat{H}}{\partial y_m}, \qquad
A_m\dot{y}_m = \frac{\partial \widehat{H}}{\partial x_m}~.
\]

In the case of a symmetrically moving vortex pair, Eqs.(11) can be
rewritten in Cartesian coordinates in the canonical form
\begin{equation}\label{eq12}
  \dot{x} = -\frac{\partial H}{\partial y}, \qquad
  \dot{y} = \frac{\partial H}{\partial x}~.
\end{equation}
with the time-independent Hamiltonian
\begin{equation}\label{eq13}
H(x, y) = \frac{A}{4\pi}\ln y - Q \, \text{arccot}\frac{x}{y}~.
\end{equation}
The trajectory of the vortex pair is parameterized by the "energy
integral" of the system (11):
\begin{equation}\label{eq14}
E = \frac{A}{4\pi}\ln y - Q \, \text{arccot}\frac{x}{y}~.
\end{equation}
Hence, the "phase" of movement along the trajectory $\Phi \equiv
\text{arccot}(x/y)$ can be expressed as
\begin{equation}\label{eq15}
\Phi = \frac{A}{4\pi Q}\ln y - \frac{E}{Q}~.
\end{equation}

Depending on vortex and flow parameters relation the case of
reverse vortex motion may be possible. The vortex pair reverses
its motion between $x_-$ and $x_+$, where
\begin{equation}\label{eq16}
x_\pm = y \frac{2\pi Q}{A}\left(1 \pm
                             \sqrt{1 -
                                  \left(\frac{A}{2\pi Q}\right)^2
                                  }
                          \right)
\end{equation}
are the turning points where $\dot{x}$ changes sign. For existence
of the turning points it has to be fulfilled the inequality
\begin{equation}\label{eq17}
(2\pi Q/A)^2 \geq 1~.
\end{equation}
If $A > 0$, then the pair moves antiparallel to the $x$ axis and
infinity corresponds the phase values
\begin{equation}\label{eq18}
\begin{split}
 \Phi = +0~, \qquad x = +\infty~, \\
 \Phi = \pi - 0~, \qquad x = -\infty~.
\end{split}
\end{equation}
If $A < 0$ the pair moves parallel to the $x$ axis. The cotangent
argument lies in this case in the interval $(-\pi,0)$ and the
phases corresponding to infinity are
\begin{equation}\label{eq19}
\begin{split}
 \Phi = -0~, \qquad x = -\infty~, \\
 \Phi = -\pi + 0~, \qquad x = +\infty~.
\end{split}
\end{equation}
It convenient to regard the vortex movement in polar coordinates,
$x = r \cos \varphi, \quad y = r \sin \varphi, \quad \varphi =
\text{arccot}(x/y)$. In this case the Eqs.(11) take the form
\begin{subequations}
\label{eq20}
\begin{equation}
\dot{r} =- \frac{A}{4\pi r}\cot\varphi +
\frac{Q}{r}~,\label{eq20-a}
\end{equation}
\begin{equation}
 \dot{\varphi} =\frac{A}{4\pi r^2}~.\label{eq20-b}
\end{equation}
\end{subequations}
These equations can also be represented in Hamiltonian form:
\begin{equation}\label{eq21}
\begin{split}
 \dot{\xi} = -\frac{\partial \tilde{H}}{\partial \varphi}~, \qquad
 \dot{\varphi} = \frac{\partial \tilde{H}}{\partial \xi}~, \\
 \tilde{H} = \frac{A}{4\pi}\ln\xi + \frac{A}{4\pi}\ln\sin^2\varphi -
 2Q\varphi~,
\end{split}
\end{equation}
where $\xi = r^2$ and Hamiltonian $\tilde{H}(\xi,\varphi)$ is
twice as large than Hamiltonian $H(x,y)$ in Cartesian coordinates
$\tilde{H}(\xi,\varphi) = 2H(x,y)$ for the same system. Thus, the
Hamiltonian equations of motion written in these coordinates are
\begin{equation}\label{eq22}
\begin{split}
 \dot{\xi} = 2Q - \frac{A}{2\pi} \cot \varphi, \\
 \dot{\varphi} = \frac{A}{4\pi\xi}.
\end{split}
\end{equation}
The distance from the center of source to the vortex reaches a
minimum $r_\ast$ when $\varphi_\ast = \text{arccot}(4\pi Q/A)$.
The equation of trajectory $r(\varphi)$ parameterized by the
energy integral is solved directly relative to the distance from
the center:
\begin{equation}\label{23}
\xi(\varphi)\equiv r^2(\varphi) = \exp \left(
             \frac{8\pi}{A} E +  \frac{8\pi Q}{A} \varphi - \ln\sin^2\varphi
                                                 \right)
\end{equation}
where $\tilde{E} = 2E$ and $E$ is given by (14). Using (23), we
integrate Eq.(20b) in quadratures:
\begin{equation}\label{24}
t - t_0 = \frac{4\pi}{A}\int_{\varphi_0}^\varphi d\varphi_1
                             \exp\left(
                             2(\mu + \lambda \varphi_1) -
                             \ln\sin^2\varphi_1
                                         \right),
\end{equation}
where
\[
\mu = \frac{4\pi E}{A}, \qquad \lambda = \frac{4\pi Q}{A}~.
\]
The function $t(\varphi)$ may be expressed in terms of elementary
functions. The inversion of this function and calculation of
$r(t)$ can be performed numerically.

Changing back to the Cartesian coordinates, we obtain the
trajectory equation
\begin{equation}\label{25}
x = y \cot \left[ \frac{A}{4\pi Q}\ln y - \frac{E}{Q} \right]~.
\end{equation}
Hence, the vortex locations at infinite distance are found by
using (19):
\begin{equation}\label{eq26}
\begin{split}
 y_{+\infty} = \exp\left( \frac{4\pi E}{A}\right), \\
 y_{-\infty} = \exp\left[ \frac{4\pi (E + \pi Q)}{A}\right].
\end{split}
\end{equation}
Thus, the vortex separation increases by a factor of $\exp(4\pi^2
Q/A)$ after the pair has passed through the flow region.

\section{BLOW-OFF OF THE VORTEX PAIR COMPONENTS}

Suppose that the energy corresponds to the desired trajectory. For
example, condition (17) must be satisfied for a trajectory with
reverse motion. Using the expression for the trajectory, we
represent (16) as
\begin{equation}\label{eq27}
x_\pm = y_\pm \, \frac{2\pi Q}{A}\left[ 1 \pm \sqrt{1 -
                                                \left(
                                                \frac{A}{4\pi Q}
                                                \right)^2}\right],
\end{equation}
where the ordinates
\begin{equation}\label{eq28}
y_\pm = \exp \left[ \frac{4\pi}{A} (E + Q\cdot
\text{arccot}\Psi_\pm)\right]
\end{equation}
are expressed in terms of the phases
\begin{equation}\label{eq29}
\Psi_\pm = \frac{2\pi Q}{A} \left[ 1 \pm\sqrt{1 -
                                                \left(
                                                \frac{A}{4\pi Q}
                                                \right)^2}\right]~.
\end{equation}
We take a value $y_0$ in the interval defined by (26) that lies
sufficiently close to a turning point. Using the expression for
the trajectory, we determine the corresponding value $x_0$,
numerically calculate the function $x(y)$, and invert it to find
the required $y(x)$.

Figure 4 demonstrates that a vortex pair components moves apart as
it approaches the source from infinity and can indeed reverse its
motion along the $x$ axis. Having traveled to a distance
sufficiently far away from the source, where the radial flow is
weaker, the pair components resume their motion in the direction
determined by the sense of vortex rotation. The separation at
negative infinity given by (26) is reached asymptotically.

\begin{figure}[bth]
\centering \includegraphics[width=7.5cm]{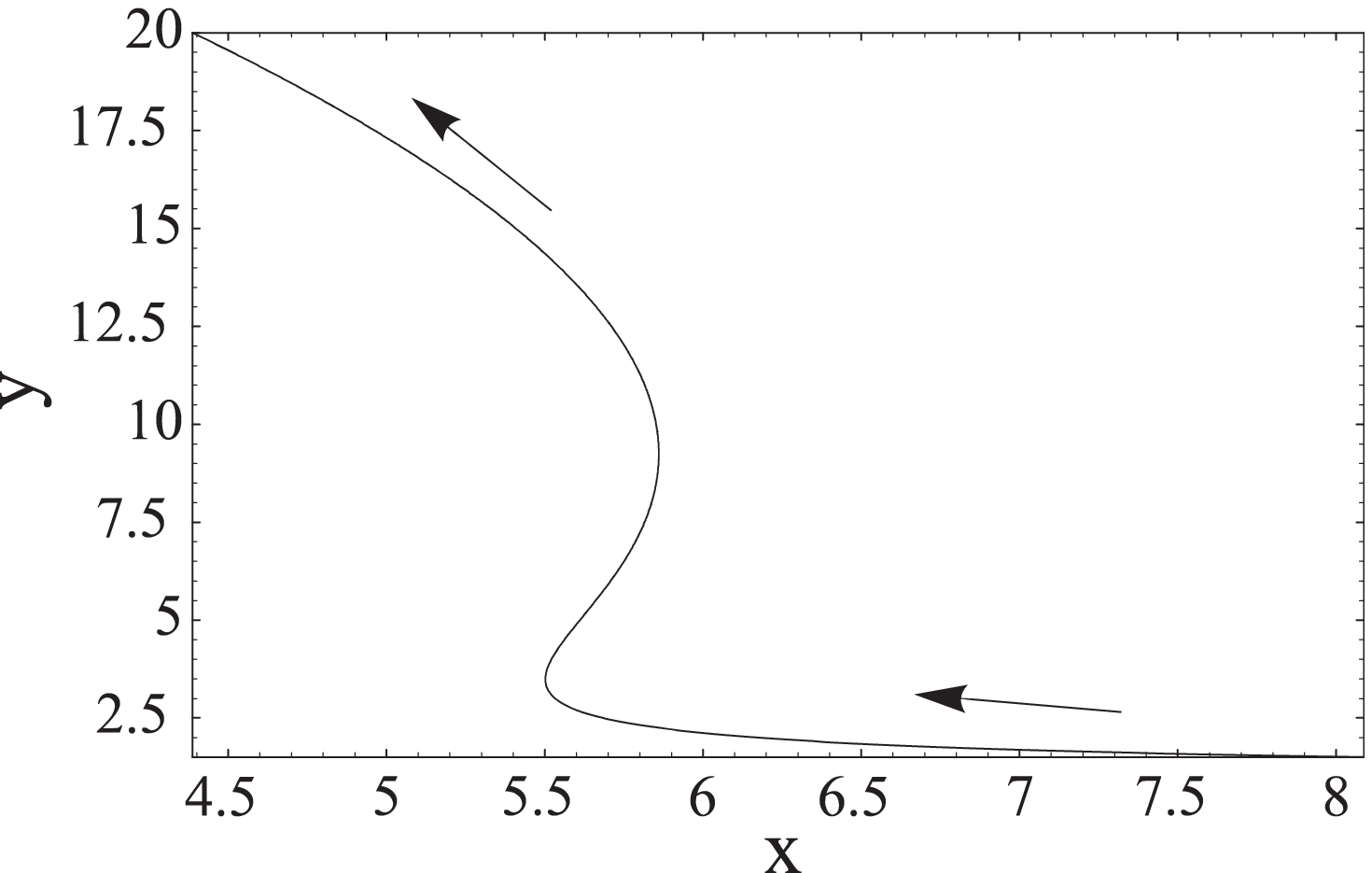} \hfill
\includegraphics
[width=7.5cm]{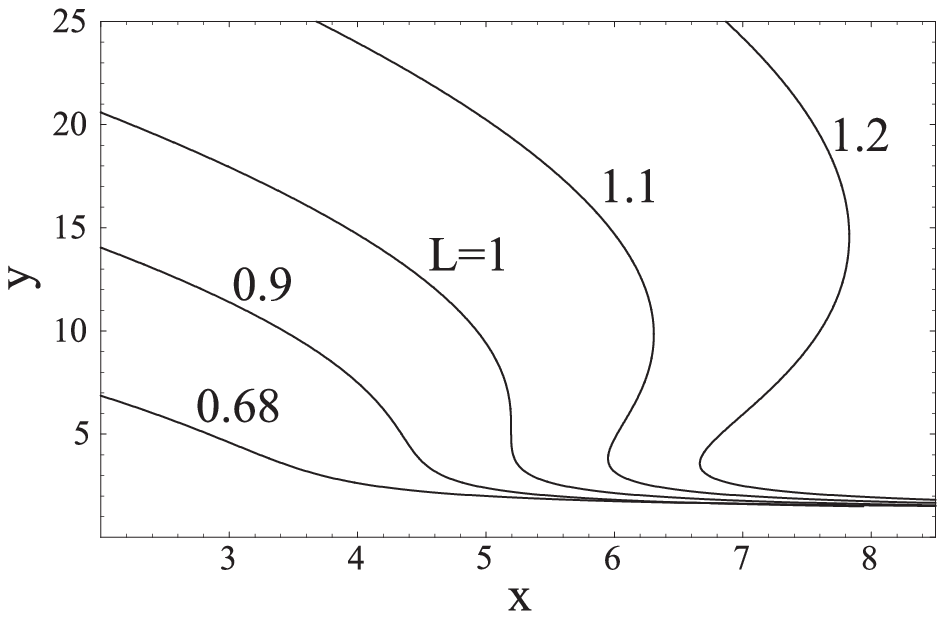}
\newline
\parbox[t]{7.5cm}
{\caption{Trajectory of a vortex in the region of reverse motion.
The arrow indicates the direction of motion of the vortex. The
parameter values $Q = 1$, \, \, $A = 2\pi - \, 0.6$, and $E = 0$
correspond to the trajectory passing through the point with $x_0 =
6.17$ and $y_0 = 2$.}\label{fig4}}
\hfill\parbox[t]{7.5cm}{\caption{Vortex trajectories for several
values of $L = 2\pi Q/A$ and $E = 0$. The curve corresponding to
$L = 1.0$ separates the regions of reverse vortex motion ($L > 1$)
and motion without turning points ($L < 1$).}\label{fig5}}
\end{figure}

Vortex trajectories in radial flow can also be calculated by
solving Eqs. (11) numerically with the values $x_0$ and $y_0$
above used as initial conditions (see caption to Fig. 4). Figure 5
shows examples of trajectories with and without reverse motion.

\section{ACCELERATION OF A VORTEX PAIR BY CONVERGING RADIAL FLOW}

Consider a vortex pair in a converging flow, as shown in Fig. 6.
\begin{figure}[bth]
\centering \includegraphics[width=7.5cm]{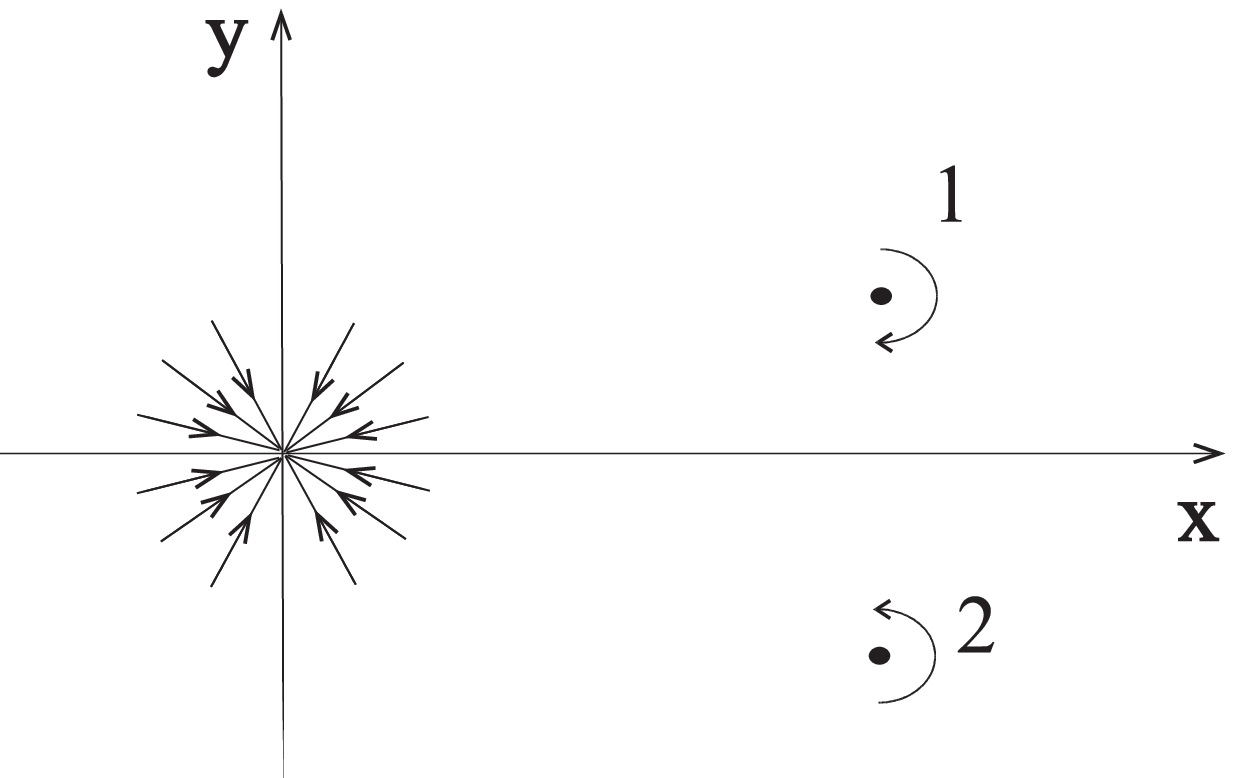} \hfill
\includegraphics
[width=7.5cm]{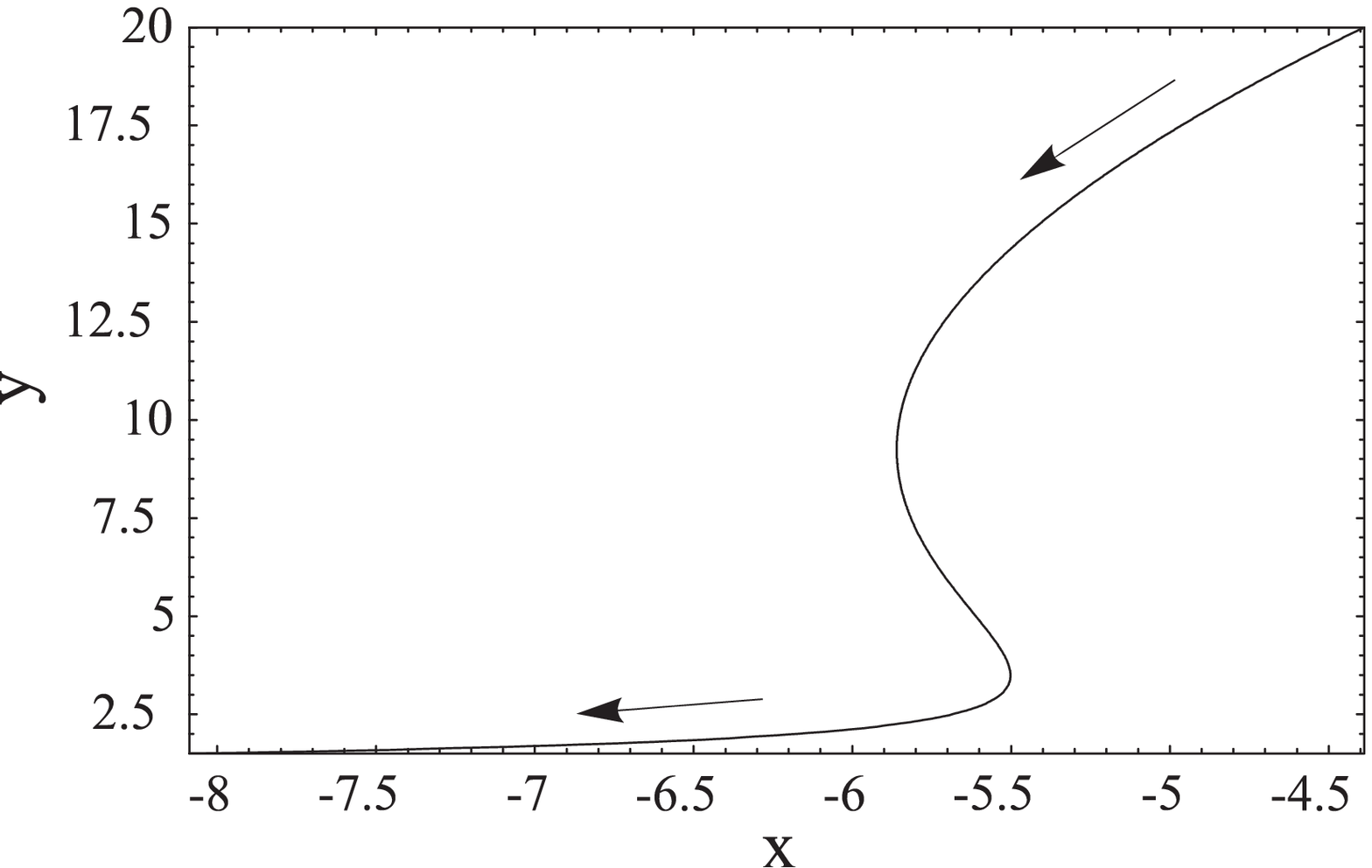}
\newline
\parbox[t]{7.5cm}
{\caption{Vortex pair in converging flow \\ (in Cartesian
coordinates).}\label{fig6}}
\hfill\parbox[t]{7.5cm}{\caption{Trajectory of a vortex pair
component in the region of reverse motion. The arrow indicates the
direction of vortex motion. The parameter values are $P = -Q = 1$,
$A = 2\pi - 0.6$, and $E = 0$.}\label{fig7}}
\end{figure}
Since $Q < 0$, we change to a new parameter according to $Q = - P
= const < 0$ and rewrite the expression for the regular
streamfunction component as
\begin{equation}\label{eq30}
\psi_r = -Q \varphi = P\varphi~.
\end{equation}
It is clear from Eq. (11b) that $\dot{y}< 0$; i.e., the vortex
separation decreases with time elapsed. According to (18), the
separation at infinite distance is
\begin{equation}\label{eq31}
y_{+\infty} = \exp\left(\frac{4\pi E}{A}\right)~, \qquad
y_{-\infty} = \exp\left[\frac{4\pi (E - \pi P)}{A}\right]~.
\end{equation}
It follows from (31) that the vortex separation decreases by a
factor of $\exp(4\pi^2 P/A)$ after the pair has passed through the
sink region.

The trajectory of the pair is found by following an analogy with
the analysis above. It is clear that the vortex components
separation decreases, and the pair can reverse its motion within a
certain portion of its trajectory near the sink if the parameters
satisfy certain conditions, as illustrated by Fig.7. Having
approached the sink to a sufficiently close distance, the pair
accelerates as the vortex separation decreases and continues to
move with a higher velocity determined by the sense of vortex
rotation after it has passed the second turning point. At negative
infinity, the pair reaches an asymptotic steady state with a
smaller vortex separation and a higher translational velocity. In
the example illustrated by Fig.7, the velocity of the pair
"thrown" out of the sink region is higher by three orders of
magnitude.

\section{LAGRANGIAN DESCRIPTION}

Hamiltonian (21) in polar coordinates can be represented as
\[
  \tilde{H} = \tilde{K}(\xi) + \tilde{U}(\varphi)~.
\]
Here, $\varphi$ and $\xi$ are interpreted as generalized
coordinate and momentum, respectively (cf.(22));

 \noindent $\tilde{K}(\xi) =
(A/4\pi)\ln \xi$ is the kinetic energy and potential energy is
expressed as
\[
\tilde{U}(\varphi) = \frac{A}{4\pi}\ln \sin^2 \varphi -
2Q\varphi~.
\]
The corresponding Lagrangian is
\begin{equation}\label{eq32}
  \tilde{L}(\varphi, \dot{\varphi}) = \frac{A}{4\pi} \ln \dot{\varphi} - \frac{A}{4\pi} \ln \sin^2
  \varphi + 2Q \varphi + \text{const}~,
\end{equation}
where the generalized velocity $\dot{\varphi} = A/4\pi \xi$ is
inversely proportional to $\xi$.\footnote{The inverse proportion
between generalized velocity $\dot{\varphi}$ and momentum $\xi$
implies that with accuracy to unessential constant summand
$\tilde{L} = \xi \dot{\varphi} - \tilde{H} = - \tilde{H}$.} Given
$\tilde{L}(\dot{\varphi}, \varphi)$, we integrate the
Euler-Lagrange equation
\begin{equation}\label{eq33}
  \frac{\ddot{\varphi}}{\dot{\varphi}^2} = \frac{4\pi}{A}
                                   \frac{\partial U(\varphi)}{\partial \varphi}
\end{equation}
to find the solution
\begin{equation}\label{eq34}
  t - t_0 = C \int_{\varphi_0}^{\varphi} \frac{d\varphi}{\sin^2\varphi}
                                          \exp \left(
                                              \frac{8 \pi Q\varphi}{A}
                                               \right)~,
\end{equation}
which is equivalent to solution (24) obtained by using the
Hamiltonian approach. In this case
\[
C = \frac{4\pi}{A}\exp \left(
                    \frac{8\pi E}{A}
                       \right)~.
\]

\section{COROTATING VORTICES IN RADIAL FLOW}

We consider two corotating vortices with equal circulations
symmetric with respect to the source of the radial flow (Fig. 8).

\begin{figure}
\includegraphics[scale=0.8]{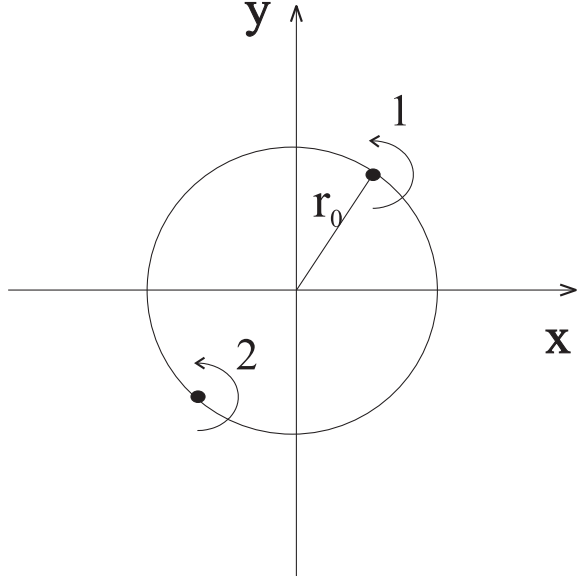}
\caption{Initial state of two corotating vortices with $A_1 = A_2
= A$. In a quiescent environment, the vortices circumrotate.}
\end{figure}

It is well known [9] that the separation $2r_0$ between the
vortices in a quiescent environment remains constant and they
rotate about the center of vorticity with an angular velocity of
$A/2\pi r_0^2$. In radial flow, equations of motion (11) for a
vortex are modified as follows:
\begin{subequations}
\label{eq35}
\begin{equation}
\dot{x} =- \frac{A}{4\pi}\frac{y}{r^2} +
Q\frac{x}{r^2}~,\label{eq35-a}
\end{equation}
\begin{equation}
 \dot{y} = \frac{A}{4\pi}\frac{x}{r^2} +
Q\frac{y}{r^2}~.\label{eq35-b}
\end{equation}
\end{subequations}
where $y = y_1 = - y_2$ and $A_1 = A_2 = A$.

System (35) yields $r^2 = 2Qt + r_0^2$; i.e., the squared
separation between vortices in a diverging flow is a linear
increasing function of time (Fig.9a). In a converging flow, the
vortex separation vanishes in the finite time $t = r_0^2/2P$
(Fig.9b), while the angular velocity of vortex rotation
indefinitely increases (see Footnote 3).

\begin{figure}
\includegraphics[scale=0.95]{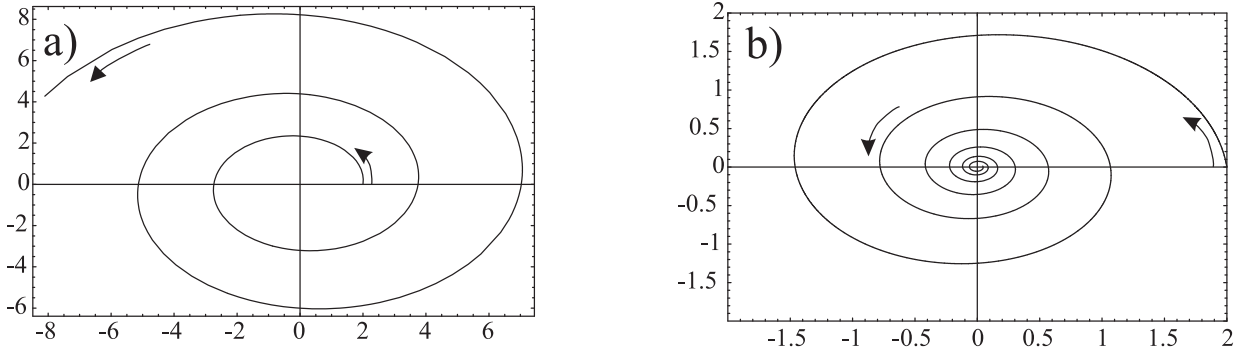}
\caption{Trajectory of a component of the pair of corotating
vortices of equal strength $A = 4\pi$: \, \, \,  (a) diverging
radial flow with $Q = 0.1$; (b) converging flow with $P =- Q =
0.1$.}
\end{figure}

Changing to the complex coordinate $w = x + i y$, we have $|w|^2 =
r^2$. Accordingly, Eqs. (35) are equivalent to
\begin{equation}\label{eq36}
  \frac{\dot{w}}{w} = \left(Q + \frac{A}{4\pi}~i\right)
  \frac{1}{|w|^2}= \left(Q + \frac{A}{4\pi}~i\right)
  \frac{1}{2Qt + r_0^2}~.
\end{equation}
Integrating Eq. (36), we obtain
\begin{equation}\label{eq37}
  w = w_0   \left(1 + \frac{2Qt}{r_0^2}\right)^
                               {\frac{1}{2}\left(1 + i \frac{A}{4\pi
                               Q}\right)}~,
\end{equation}
where $w_0$ is the initial value of the complex coordinate. In the
limit of $Q \rightarrow 0$, the solution describes uniform
rotation. Indeed, changing to the new variable $\eta = 2Qt/r_0^2$,
we have
\begin{equation}\label{eq38}
  w^2 = w_0^2 \cdot [(1 + \eta)^{1/\eta}]^{i At/2\pi r_0^2}~.
\end{equation}
Since $\underset{\eta \rightarrow 0}{\lim} (1 + \eta)^{1/\eta} =
e$, we obtain
\[
w \rightarrow w_0 \exp \left(i \frac{At}{2\pi r_0^2}\right)~.
\]
In the polar coordinates, the motion is particularly simple. It is
obvious that the instantaneous angular velocity is expressed as
follows ($\tilde{H} = A \ln \xi/4\pi - 2Q\varphi)$:
\[
\dot{\varphi} (t) = \frac{A}{4\pi (2Qt + r_0^2)}~.
\]
Hence,
\[
\varphi(t) = \frac{A}{8\pi Q}\ln \left(1 +
\frac{2Q}{r_0^2}t\right) + \varphi(0)~;
\]
i.e., the trajectory is a logarithmic spiral. The limit as $Q
\rightarrow 0$ is easily taken.

\section{VORTICES IN ROTATING RADIAL FLOW}

Flow rotation can be taken into account by adding a corresponding
term to the expression for the regular streamfunction component,
\[
\psi_r = -Q\varphi + \psi_r^\Omega~,
\]
where $\psi_r^\Omega = \Omega r^2/2$ represents rigid-body
rotation with angular velocity $\Omega$. Then, Eqs. (10) are
rewritten in terms of $w = x + i y$ as
\begin{equation}\label{eq39}
\begin{split}
\frac{\partial w_1}{\partial t} - i\Omega w_1 = \frac{i A_2}{2\pi
r_{12}^2}(w_1 - w_2) + \frac{Q}{r_1^2}w_1~,
\\ \frac{\partial w_2}{\partial t} - i\Omega w_2 = - \frac{i A_1}{2\pi
r_{12}^2}(w_1 - w_2) + \frac{Q}{r_2^2}w_2~.
\end{split}
\end{equation}
In the rotating reference frame we substitute $w = w'\exp(i\Omega
t)$ and use the relations $r^2 = |w|^2 = |w'|^2$ and $r_{12}^2 =
|w_1 - w_2|^2 = |w'_1 - w'_2|^2$ to derive the equations of motion
for vortices in irrotational radial flow, which makes it possible
to use the results obtained above.

\section{CONCLUSIONS}

We have shown that the problem of vortex pair motion in
two-dimensional radial flow has an exact solution. Under certain
conditions for flow parameters, the vortex pair can reverse its
motion. In a diverging flow, the separation between the vortices
increases, but the pair does not break apart. In a converging
flow, the vortices move inwards, and the translational velocity of
the pair increases. In both cases, the vortex separation in the
pair changes by a finite amount as it passes through the source
region. The squared separation between corotating vortices of
equal strength is a linear function of time. In a diverging flow,
they move apart indefinitely. In a converging flow, the vortex
separation vanishes in a finite time. Note that the problem of
vortex motion in rotating radial flow may have qualitative
applications in planetary geophysics. For example, the recently
discovered double-lobed vortex swirling around Venus' south pole
[12] may be described by a similar model if there is meridional
flow in the polar region. The present model cannot describe
vortices spun up by the wind, since circulation is conserved.
Varying circulation can be treated by allowing for dissipation. It
would be interesting to analyze the asymmetrical motion of a
vortex pair in a radial flow. Finally, also of interest is the
motion of vortex rings in radial flow. \hspace {2cm}
\it{Translated by A.~Betev}

\end{document}